\documentclass[aps,prl,twocolumn,superscriptaddress,showpacs,floatfix]{revtex4-2}

\usepackage{amsmath,amssymb,amsfonts,float,graphics,epsfig,epstopdf,color,verbatim,tabularx,bm,multirow,appendix}
\usepackage[utf8]{inputenc}
\usepackage[T1]{fontenc}
\usepackage{xcolor}

\usepackage{dsfont}
\usepackage{textcomp}
\usepackage{yfonts}
\usepackage{bm}
\usepackage{subfigure}
\usepackage{mathrsfs}
\usepackage{graphicx}
\usepackage{verbatim}
\usepackage{hyperref}
\usepackage{multirow}
\usepackage{braket}
\usepackage[normalem]{ulem}
\usepackage{tikz}
\usetikzlibrary{calc}
\usetikzlibrary{shapes.multipart}

\renewcommand{\vec}[1]{{\mathbf #1}}

\newcommand{\comments}[1]{}

\newcommand{\stkout}[1]{\ifmmode\text{\sout{\ensuremath{#1}}}\else\sout{#1}\fi}

\makeatletter
\def\l@subsubsection#1#2{}
\makeatother

\begin{document}

\title{Different temperature-dependence for the edge and bulk of entanglement Hamiltonian}

\author{Menghan Song}
\affiliation{Department of Physics and HKU-UCAS Joint Institute of Theoretical and Computational Physics, The University of Hong Kong, Pokfulam Road, Hong Kong SAR, China}

\author{Jiarui Zhao}
\affiliation{Department of Physics and HKU-UCAS Joint Institute of Theoretical and Computational Physics, The University of Hong Kong, Pokfulam Road, Hong Kong SAR, China}

\author{Zheng Yan}
\email{zhengyan@westlake.edu.cn}
\affiliation{Department of Physics, School of Science, Westlake University, 600 Dunyu Road, Hangzhou 310030, Zhejiang Province, China}
\affiliation{Institute of Natural Sciences, Westlake Institute for Advanced Study, 18 Shilongshan Road, Hangzhou 310024, Zhejiang Province, China}
\affiliation{Department of Physics and HKU-UCAS Joint Institute of Theoretical and Computational Physics, The University of Hong Kong, Pokfulam Road, Hong Kong SAR, China}

\author{Zi Yang Meng}
\affiliation{Department of Physics and HKU-UCAS Joint Institute of Theoretical and Computational Physics, The University of Hong Kong, Pokfulam Road, Hong Kong SAR, China}

\begin{abstract}
We propose a physical picture based on the wormhole effect of the path-integral formulation to explain the mechanism of entanglement spectrum (ES), such that, our picture not only explains the topological state with bulk-edge correspondence of the energy spectrum and ES (the Li and Haldane conjecture), but is generically applicable to other systems independent of their topological properties. We point out it is ultimately the relative strength of bulk energy gap (multiplied with inverse temperature $\beta=1/T$) with respect to the edge energy gap that determines the behavior of the low-lying ES of the system. Depending on the circumstances, the ES can resemble the energy spectrum of the virtual edge, but can also represent that of the virtual bulk. \textcolor{black}{We design models both in 1D and 2D to successfully demonstrate the bulk-like low-lying ES at finite temperatures, in addition to the edge-like case conjectured by Li and Haldane at zero temperature.} Our results support the generality of viewing the ES as the wormhole effect in the path integral and the different temperature-dependence for the edge and bulk of ES.
\end{abstract}

\date{\today}
\maketitle

\noindent{\textcolor{blue}{\it Introduction.}---} More than a decade ago, Li and Haldane~\cite{Li2008entangle} proposed that the entanglement spectrum (ES) is a direct measurement of the topological properties of quantum many-body systems, in that the low-lying ES are closely related to the true energy spectra on the edges of open boundary systems. \textcolor{black}{Entanglement entropy (EE) as another important measure of quantum correlation has been studied carefully over the years. It is pointed out that EE at finite temperatures obeys a volume law and resembles the thermal dynamic entropy of small subsystems~\cite{Korepin2004,Nakagawa2018,Kitaev2006}.} Although the generality of Li and Haldane's statement has been questioned~\cite{chandranHow2014}, it is still believed by many that overall the ES reveals more entanglement information  and other non-local observables~\cite{Thomale2010nonlocal,Poilblanc2010entanglement,Calabrese_2004,Fradkin2006entangle,Pollmann2010entangle,FSong2012,Klich2009,Song2011,Gioev2006,jiangFermion2022,JRZhao2020,BBChen2022,wangScaling2021,Luitz2014}. 
Later, Qi, Katsura and Ludwig analytically demonstrated the bulk-edge correspondence between the ES of (2+1)D gapped topological states and the energy spectrum on their (1+1)D edges~\cite{XLQi2012}. However, apart from these gapped topological states, the universality of Li and Haldane conjecture still remains an open and intriguing question. \textcolor{black} {Previous works on EE~\cite{Korepin2004,Nakagawa2018,Kitaev2006} and reduced density matrix~\cite{Popescu2006} suggest the resemblance between the bulk spectra of subsystem and the entanglement spectra at sufficiently high temperatures, while the relationship between the edge mode and bulk mode is somehow unknown so far.} What's more, besides the abstract mathematical proofs, there exists no intuitive and transparent physical picture that could penetrate the barrier between the microscopic lattice models and the field-theoretical continuum description and to explain these phenomena in simple language.

Part of the reason, that the Li and Haldane conjecture still remains a conjecture is because it is very hard to compute the ES for generic 2D or 3D quantum many-body systems. Due to the exponentially growth of computation complexity and memory cost~\cite{Poilblanc2010entanglement,Thomale2010nonlocal,Pollmann2010entangle,ciracEntanglement2011,fidkowskiEntanglement2010,weizhu2019reconstrcting,weizhu2020entanglement,Tang2020critical}, systems with long boundaries of entanglement region and at higher dimensions are in general prohibited. Besides few exactly solvable limits, most of the ES studies by exact diagonalization (ED) and density matrix renormalization group (DMRG) so far have focused on (quasi) 1D \textcolor{black}{quantum} systems. \textcolor{black}{For 2D, several ED/DMRG studies have extracted the ES of systems with small width~\cite{Popescu2006,Alba2013,Kolley2013}. In fact, many of the ES studies target topological phases~\cite{Reg2009,Bray2009,Fidkowski2010,Zhu2015,Thomale2010,Pollmann2010entangle,Kargarian2010,Fidkowski2011,Schliemann2011,Yao2010}. Methods that can probe generic ES information in more common systems at high dimensions and with larger sizes are still in demand.}

Quantum Monte Carlo (QMC), on the other hand, usually stands out as the powerful tool to explore quantum many-body systems with larger sizes and at higher dimensions, as the importance sampling scheme can in principle reduce the exponential complexity into a polynomial one~\cite{sandvikComputational2010,assaadWorld2008,Sandvik2019SSE,xuRevealing2019,XuZhang2021,panSign2022,panSport2022}. Although QMC in a path-integral formulation accesses the partition function instead of the ground state wave function, it has been shown that the computation of Rényi EE can be achieved by sampling the partition function in a replicated manifold with different boundary conditions for the entanglement region $A$ and the environment $\overline{A}$ of the system~\cite{CARDY1988,Calabrese_2004,Hastings2010,Humeniuk2012,Inglis2013,InglisNJP2013,KallinPRL2013}. {\color{black}Consequently, the entanglement signature and scaling behavior of many novel phases and phase transitions in EE have been reliably extracted in QMC simulations in higher dimension systems~\cite{JRZhao2020,JRZhao2021,Luitz2014,KallinJS2014,Helmes2014,DEmidio2020,Guo_2021,chenTopological2022,jiangFermion2022,demidioUniversal2022,liuFermion2022,liaoTeaching2023,panComputing2023,songDeconfined2023}.}

\begin{figure}[htp!]
	\centering
	\includegraphics[width=\columnwidth]{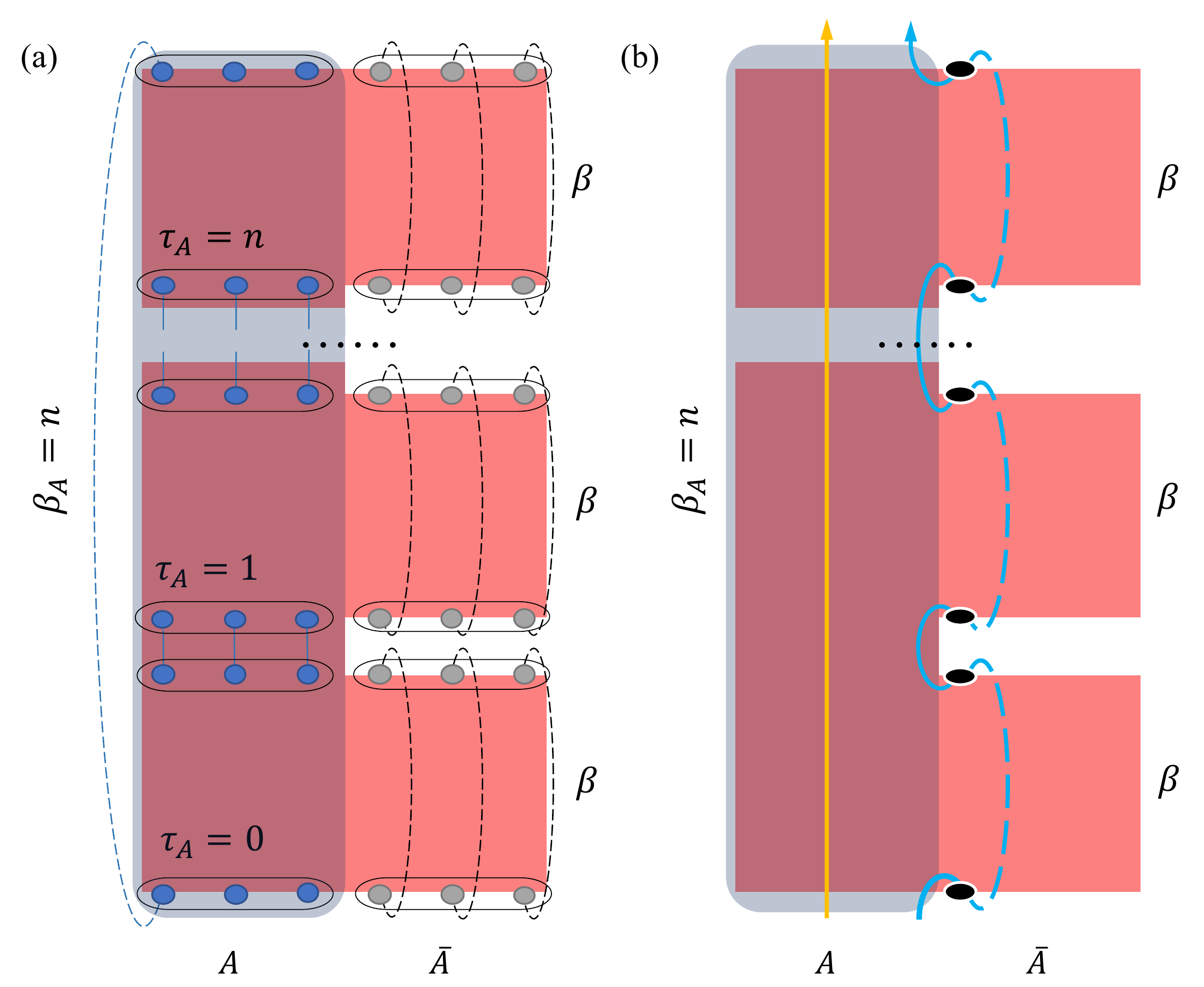}
	\caption{(a) Graphical representation of the partition function $\mathcal{Z}_{A}^{(n)}$ in the replicated manifold. The shaded area is the entanglement region ${A}$ in which all the replicas are glued together along the imaginary time and has length $\beta_A=n\beta$. In the environment region $\bar{A}$, replicas are independent along the imaginary time axis and each has length $\beta$. (b) Two different worldline paths along the imaginary time. Yellow one travels inside the bulk while blue one goes into the environment and experiences the wormhole effect. Black circles are the wormholes which teleport a worldline to the other side of a replica in $\bar{A}$ via the virtual path (blue dashed line).}
	\label{fig:fig1}
\end{figure}

\noindent{\textcolor{blue}{\it The wormhole picture of ES.}---} Recently, some of us extended the QMC computation of the EE to that of the ES and have successfully reduced the computational complexity and made the computation of ES with long entanglement boundaries and in higher dimensions possible~\cite{yanRelating2022,Wu2023classical}. The basic idea is that in the replicated manifold with partition function (as shown in Fig.~\ref{fig:fig1} (a))
\begin{equation}
\mathcal{Z}_A^{(n)}=\operatorname{Tr}\left[\rho_A^n\right]=\operatorname{Tr}\left[e^{-n \mathcal{H}_A}\right],
\label{eq:eq1}
\end{equation}
where $\rho_A=\mathrm{Tr}_{\overline{A}}\rho$ is the reduced density matrix (RDM), defined as the partial trace of the total density matrix $\rho$ of Hamiltonian $H$ over a complete basis of $\overline{A}$ and $\mathcal{H}_A$ is the corresponding entanglement Hamiltonian (EH), one can define the effective imaginary time $\beta_{A}=n$ for the EH at integer points $n =1,2,3,...$. It's worthwhile to note that the $\beta_A$ shall be integer points in the QMC simulation because we can only simulate the whole RDM $\rho_A=\mathrm{Tr}_{\overline{A}}(e^{-\beta H})$ instead of a fractional one. One can then compute the dynamic correlation functions $G(\tau_A)$ at these integer time points. The ES can be readily obtained from the imaginary time correlations of EH via numeric analytic continuation methods, such as stochastic analytic continuation (SAC)~\cite{sandvik1998stochastic,Beach2004,OFS2008using,sandvik2016constrained,Zhou2021amplitude,ZY2020,YCWang2021vestigial,YCWang2020,jiangMonte2022,zhouEvolution2022}.

The physical picture of the above process of computing the ES in the path-integral is described clearly via the schematic diagrams in Fig.~\ref{fig:fig1}, where Fig.~\ref{fig:fig1} (a) is the replicated manifold of the path-integral of $\mathcal{Z}_A^{(n)}$ in Eq.~\eqref{eq:eq1}, and a wormhole effect emerges at the entangled edges in Fig.~\ref{fig:fig1} (b). \textcolor{black}{During QMC simulation}, there are two typical paths of the worldline~\cite{Syljuaasen2002,sandvikComputational2010,Sandvik1991,Sandvik1999,Syljuaasen2002,ZY2019,yan2020improved} in the path-integral in the Fig.~\ref{fig:fig1} (b): the yellow one deep in the bulk goes through all the replicas with an imaginary time length $\sim n\beta$; the blue one at the entangled edge can take a much shorter path because of the periodic boundary condition (PBC) of length  $\beta$ of every replica in $\overline{A}$. The dashed blue line shows how the wormholes (the black circles in Fig.~\ref{fig:fig1} (b)) teleport the worldline from the bottom to top of one replica without propagating through $\beta$ due to such PBC, so that the total imaginary time length of the wormhole path is only about $\sim n$.  We therefore dubbed this shorter path effect cased by the different connections of the $\mathcal{Z}_A^{(n)}$ in the space-time as the "wormhole" picutre of ES~\cite{yanRelating2022}.

In the path-integral, the shorter path always leads to more important contribution. The spectral function $S(\omega)$ for physical observable, represented as $\mathcal{O}$, can be written by the eigenstates $|n\rangle$ with the eigenvalue $E_n$ of the Hamiltonian $\mathcal{H}$,
$S(\omega)=\frac1\pi\sum_{m,n}e^{-\beta E_n}|\langle m|\mathcal{O} |n\rangle|^2\delta(\omega-[E_m-E_n])$. Therefore, there is a relation between energy spectrum $S(\omega)$ and imaginary time correlation $G(\tau)$ as $G(\tau)=\int_0^\infty d\omega K(\omega,\tau) S(\omega)$. The $K(\omega,\tau)$ is a kernel with slightly different expressions for bosonic/fermionic $\mathcal{O}$. Obviously, the imaginary time correlation $G(\tau)$ is the summation of all the gap modes. The important modes have larger weights in the summation. When the temperature is low enough ($\beta=\infty$), the $S(\omega)$ can be treated as $G(\tau)\sim e^{-\tau \Delta}$, where the $\Delta$ is the first-excited gap and $\tau$ is the length of imaginary time. In our wormhole picture, the time length is different for edge and bulk, thus the wormhole path plays the important role.
As pointed out in Ref.~\cite{yanRelating2022}, we can simply estimate the ratio of the exponential factors deep inside the bulk and that at the edge to be roughly $\beta\Delta_b : \Delta_e$. At the ground state, $\beta\rightarrow\infty$ renders $\beta\Delta_b \gg \Delta_e$ and edge path will dominate over the path-integeral and be more important for the low-energy ES, which is the Li and Haldane conjucture. But one immediately sees from here that the Li and Haldane conjecture is not only limited to the topological phase (gapped bulk and gapless edge), but also works in the cases where both bulk and edge are gapped~\cite{liu2023probing}. Moreover, since finite size systems always acquire finite size gaps, the low-lying ES can resemble the edge spectra when the temperature is low enough, but for systems at finite temperatures, when the edge exponential factor is much larger than the bulk one, and the low-lying ES will resemble the bulk energy spectra. \textcolor{black}{It is surprising that one can also obtain the bulk-like low-lying ES at finite temperature, hinted in Ref.~\cite{yanRelating2022}, and the wormhole mechanism offers a direct and lucid picture to understand the appearance of bulk information at finite temperature implied in previous works~\cite{Korepin2004,Nakagawa2018,Kitaev2006,Popescu2006}. In this article, we focus on systematic demonstration of such wormhole picture and the different temperature-dependence for the edge and bulk of entanglement Hamiltonian via QMC simulations in 1D and 2D quantum spin systems.}

\begin{figure}[htp!]
	\centering
	\includegraphics[width=\columnwidth]{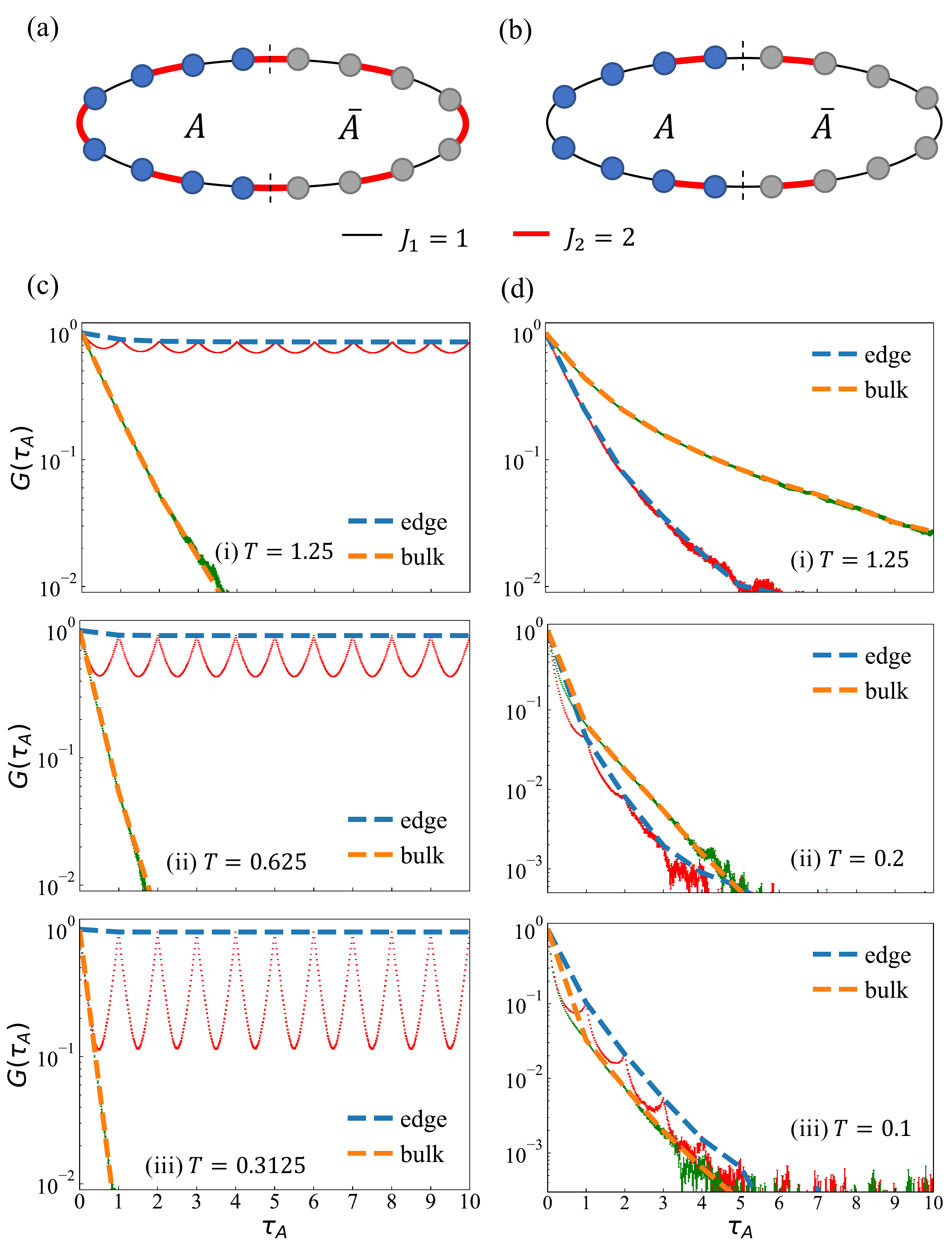}
	\caption{Schematic lattice models and ES results for 1D Heisenberg chains. (a) 1D chain with dimers inside the bulk. (b) 1D chain with stronger bonds at entanglement boundaries. (c) Decay of correlation function along the imaginary time for the bulk-gapped chain as in (a). (d) Decay of correlation function along the imaginary time for the edge-gapped chain as in (b). Sub-figures (i), (ii) and (iii) correspond to cases with decreasing temperature $T$ (increasing $\beta$). }
	\label{fig:fig2}
\end{figure}

\noindent{\textcolor{blue}{\it 1D Heisenberg Chains.}---}
Here we employ the stochastic series expansion QMC method~\cite{sandvikComputational2010,Sandvik1991,Sandvik1999,Syljuaasen2002,ZY2019,yan2020improved} to simulate the RDM of the entanglement region $A$ via a replicated manifold of 1D antiferromagnetic (AFM) Heisenberg spin chains. The lattice models are shown in Fig.~\ref{fig:fig2} (a) and (b), with varying coupling strength on the edge and in the bluk,
\begin{equation}
H=\sum_{\left \langle i,j \right \rangle }J_{i,j}\mathbf{S}_{i}\cdot \mathbf{S}_{j}
\end{equation}
where $\left \langle i,j \right \rangle $ denotes nearest neighbors and the strength of each bond $J_{i,j}$ is chosen to tune the gaps of $\Delta_{b/e}$ so that the wormhole effects can be demonstrated prominently by the correlation functions in the bulk and on the edge of $A$.

From the partition function of Eq.~\eqref{eq:eq1}, we measure the imaginary time correlation function
\begin{equation}
G(\tau_A) = \langle \vec{S}_i(\tau_A)\cdot \vec{S}_j(0)\rangle = \frac{\operatorname{Tr}[e^{-(n-\tau_A)\mathcal{H}_A}\vec{S_i}e^{-\tau_A\mathcal{H}_A}\vec{S_j}]}{\operatorname{Tr}[e^{-n\mathcal{H}_A}]}
\label{eq:eq2}
\end{equation}
where $\tau_{A}\in [0,\beta_A]$ and must be an integer. In the spectral representation, the long time decay of the bulk/edge correlation function becomes $G(\tau_{A})\sim e^{-\Delta_{b/e}  \tau_{A}}$, where $\Delta_{b/e}$ are the lowest energy gaps of bulk/edge in the ES. Therefore, a larger gap results in a faster decay of $G(\tau_{A})$ along the imaginary time. By observing the decay rate of the correlation function at both entanglement edges ($i,j \in \partial A$ or $i,j \in \partial \overline{A}$) and in the bulk ($i,j \in A$), we identify the relative gap size between edge and bulk. Since we focus on the Heisenberg spin model with $SU(2)$ symmetry, all the correlation can be further simplified in the $S^{z}$ basis on the same site, i.e., $G(\tau_{A})=\left \langle S^{z}_i(\tau_{A})S^{z}_i(0) \right \rangle$.

We have also calculated the correlation functions inside each replica measured at continuous imaginary time to demonstrate the wormhole effect. We note that the time inside replicas is not the imaginary time for the evolution of $\mathcal{H}_A$, but that of the original Hamiltonian $H$, which means the entanglement Hamiltonian experience different effective temperature than $1/\beta$. In Fig.~\ref{fig:fig2} (c) and (d), the dotted lines represent the correlations measured at continuous time points while the dashed lines are those measured at integer time points $\tau_{A}=1,2,...,n$ which contribute to the ES. It is interesting to see that the continuous time correlation functions on edges have valleys (red dotted line) between the two integer $\tau_A$ points, whereas those of the bulk spins (green dotted line) coincide with the correlations measured at integer time points (orange dashed line). Such phenomena exactly reflect the working of the wormhole effect, in that, the worldlines at edge sites can easily go into the environment $\bar{A}$ and reach next $\tau_{A}$ with fewer attenuation compared with a path inside the bulk of $A$.


We simulate 1D $L = 32$ chain with spatial PBC and different bond settings as schematically shown in Fig.~\ref{fig:fig2} (a) and (b). At first, we reproduce the normal Li and Haldane conjecture from a dimmerized chain with bond strength $J_1=1$ and $J_2=2$, as shown in Fig.~\ref{fig:fig2} (a). The dimerized bulk state favors a gapless edge state when the strong bonds are cut at the boundary. According to the conjecture, ES in this case would resemble the gapless edge mode of the energy spectrum. Then we measure the imaginary time correlations $G(\tau_{A})$ on the replicated manifold. In Fig.~\ref{fig:fig2}(c) (i), at $T=1.25$, the correlation functions of bulk sites (orange dashed line) decay rapidly along the imaginary time, which manifests a large energy gap $\Delta_b$ inside the bulk according to $G(\tau_A)\sim e^{-\Delta_b \tau_A}$. On the other hand, the blue dashed line which is the correlations of edge spins, decays very slowly, indicating a gapless mode. It means the low-lying mode of the ES is from the entangled edges, which is highly consistent with the Li and Haldane conjecture.

As discussed above, we present two decaying modes, one is measured along continuous time points and another is obtained at $\tau_{A}=1,2,...,n$. When decreasing temperature to $T=0.625$ and $T=0.3125$ in Fig.~\ref{fig:fig2} (c) (ii) and (iii) to approach the ground state, the continuous time correlations in the bulk tend to decay faster, however the continuous time correlations on the edge bounce back at integer $\tau_{A}$ points to meet the $G(\tau_A)$ and leads to a even flatter decaying mode on the edge at $\tau_{A}=1,2,...,n$, manifesting a gapless ES. Therefore, an exactly gapless edge mode in this case is guaranteed at ground state $\beta \to \infty$ by the wormhole effect.

\textcolor{black}{More interestingly, according to our path-integral picture, the low-lying ES can represent the bulk mode when $\beta\Delta_b<\Delta_e$.} We design such situation in Fig.~\ref{fig:fig2} (b) whose bulk is a normal AFM Heisenberg chain with gapless mode~\cite{PhysRev.128.2131,Poilblanc2010entanglement}, but utilize stronger bonds near the entanglement boundaries to gap out the edge spins. The obtained correlation functions are shown in Fig.~\ref{fig:fig2}(d). At low temperature ($T=0.1$ in Fig.~\ref{fig:fig2}(d) (iii)) the correlation of edge spins decays slower than that of bulk spins, again consistent with the Li and Haldane as $\beta\Delta_b >  \Delta_e$. But when we increase the temperature $T$ (decrease $\beta$) to weaken the wormhole effect, the reversal of $\beta\Delta_b < \Delta_e$ happens and our data clearly exhibit the bulk-like ES. In Fig.~\ref{fig:fig2}(d) (i) at $T=1.25$ and (ii) at $T=0.2$, we can see that the edge mode decays faster than the bulk mode and therefore becomes much more gapped at larger $T$ (smaller $\beta$). \textcolor{black}{Therefore, ES resemble the bulk energy spectra at high enough temperatures.}

\begin{figure}[htp!]
	\centering
	\includegraphics[width=\columnwidth]{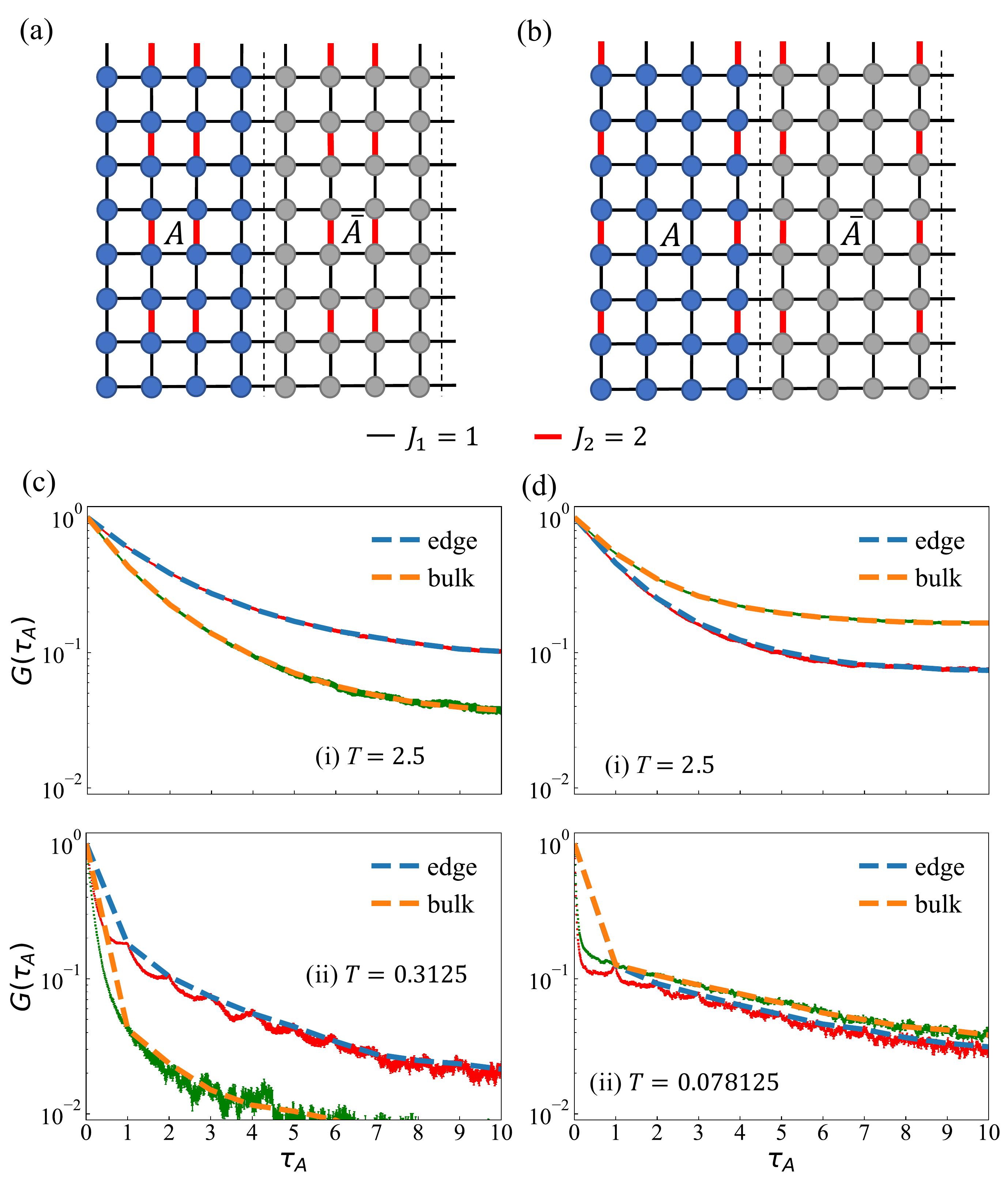}
	\caption{Schematic lattice model and ES results for 2D Heisenberg models. (a) 2D square lattice with stronger bonds $J_2=2$ inside the bulk and weak bonds $J_1=1$ at the entanglement boundaries. (b) 2D square lattice with stronger bonds at the entanglement boundaries but weak bonds inside the bulk. (c) Decay of correlation function along the imaginary time for the bulk-gapped model as in (a). (d) Decay of correlation function along the imaginary time for the edge-gapped model as in (b). Sub-figures (i) and (ii) correspond to cases with decreasing temperature $T$ (increasing $\beta$). }
	\label{fig:fig3}
\end{figure}

\noindent{\textcolor{blue}{\it 2D Heisenberg Model.}---}
\textcolor{black}{The picture based on the wormhole effect in the path-integral is independent of the details of the models and dimensions. The different temperature-dependence of edge and bulk EH can also be demonstrated in 2D systems.} To this end, we design a two-dimensional AFM Heisenberg model on square lattice with different coupling bonds, as shown in Fig.~\ref{fig:fig3} (a) and (b). The linear system size is $L=32$ and the simulation geometry of the lattice is a torus with the entanglement region $A$ and the environment $\overline{A}$ offering the bipartition.

Similar as in the 1D case, we first reproduce the normal Li and Haldane conjecture with the setting in Fig.~\ref{fig:fig3} (a). The stronger bonds $J_2=2$ generate dimers in the bulk such that the bulk is gapped and virtual edge with $J_1=1$ is gapless, and the ES must be like the edge modes. The correlation functions are shown in Fig.~\ref{fig:fig3} (c). At $T=2.5$ (Fig.~\ref{fig:fig3} (c) (i)), the imaginary time correlation $G(\tau_{A})$ of EH at edge decays much slower than that in the bulk and it is even more so at lower temperature $T=0.3125$ (Fig.~\ref{fig:fig3} (c) (ii)). The bounce of the continuous time correlation on the edge (the red dotted line) to meet the $G(\tau_A)$ at integer time points (the blue dashed line) is also seen at lower temperature.

For the case of Fig.~\ref{fig:fig3} (b), the bulk is a normal AFM Heisenberg model which has gapless Goldstone mode at the ground state~\cite{Anderson1952,Oguchi1960,Shao2017nearly,chakravartyTwo1989}, while the edge is gapped by stronger bonds. \textcolor{black}{As shown in Fig.~\ref{fig:fig3} (d) (ii), the Li and Haldane conjecture is still tenable at low temperature $T=0.078125$ because of the wormhole effect, although the long time $G(\tau_A)$ on edge (the blue dashed line) has a lower value compared with that in the bulk (the orange dashend line), suggesting a smaller spectral weight in the ES due to the setting of the gapped edge. More importantly, similar to 1D case, the role of bulk and edge interchanges as we increase temperature $T$ in Fig.~\ref{fig:fig3} (d) (i) with $T=2.5$ to weaken the wormhole effect. Here one sees the bulk $G(\tau_A)$ (the orange dashed line) decay much slower than the edge $G(\tau_A)$ (the blue dashed line), and consequently the bulk-like low-lying ES manifest.}

With the 1D and 2D examples, we have demonstrated the working of our wormhole picture based on the path-integral of replicated manifold. These results strongly support the generality of our picture, in any dimension, and the Li and Haldane conjecture is one limiting case, $\beta \to \infty$ in our framework.

\noindent{\textcolor{blue}{\it Discussion.}---}
It is well accepted that the understanding of the information encoded in the ES, especially for interacting systems at 2D or higher dimensions, is of fundamental value and still far from complete. In this context, our work points out that it is the wormhole effect of the path-integral on replicated manifold that could unlock the mechanism of ES. The famous Li and Haldane conjecture is one limiting case in our general physical picture, in that, the low-lying ES will always be like the edge energy spectra at low temperatures, independent of whether the system is topological or not. \textcolor{black}{What's more, via engineering the coupling strength and temperature according to our physical picture, a bulk-like ES can be easily realized. Guided by such generic picture and the computation scheme presented in this work, it is expected that desired entanglement properties for 2D or higher dimensional quantum many-body systems, can be accessed to broaden our understanding of ES.}

{\it{Acknowledgment.-}} We thank Bin-Bin Chen, Xiyue Lin, Han Li, Tao Shi, Wei Li, Fabien Alet and Subir Sachdev for valuable discussions on related topics. ZY also thanks Shangqiang Ning, Zenan Liu, Rui-Zhen Huang for helpful discussions. We cknowledge support from the RGC of Hong Kong SAR of China (Project Nos. 17301420, 17301721, AoE/P-701/20, 17309822,  HKU C7037-22G), the ANR/RGC Joint Research Scheme sponsored by Research Grants Council of Hong Kong SAR of China and French National Research Agency(Project No. A\_HKU703/22), and the Seed Funding ``Quantum-Inspired explainable-AI'' at the HKU-TCL Joint Research Centre for Artificial Intelligence. We thank the HPC2021 system under the Information Technology Services and the Blackbody HPC system at the Department of Physics, the University of Hong Kong for their technical support and generous allocation of CPU time.

\bibliography{wormhole}
\end{document}